\documentclass[twocolumn]{autart}
\usepackage{mathrsfs}
\usepackage{amsfonts}
\usepackage{amsmath,amssymb}
\usepackage{graphicx}
\usepackage{float}
\usepackage[dvips]{epsfig}
\usepackage{color}

\def\dref#1{(\ref{#1})}

\begin{document}

\begin{frontmatter}
%\runtitle{Insert a suggested running title}  % Running title for regular
                                              % papers but only if the title
                                              % is over 5 words. Running title
                                              % is not shown in output.

\title{Robust Consensus for Multi-Agent Systems Communicating over Stochastic  Uncertain Networks \thanksref{footnoteinfo}} % Title, preferably not more
                                                % than 10 words.

\thanks[footnoteinfo]{This work was supported in part by the National Natural Science Foundation of China under grants 61473005, 11332001, 61225013,
and in part by the Hong Kong RGC under the project CityU 111613,
CityU 11200415.}

\author[China]{Zhongkui Li}\ead{zhongkli@pku.edu.cn},
\author[UK]{Jie Chen}\ead{jichen@cityu.edu.hk}

         % full addresses

\address[China]{State Key Laboratory for Turbulence and Complex Systems, Department of Mechanics and Engineering Science,
College of Engineering, Peking University, Beijing 100871, China}

\address[UK]{Department of Electronic Engineering, City University of Hong Kong, Kowloon, Hong Kong}

\begin{keyword}
Cooperative control, consensus, uncertainty, communication channel, robustness.
\end{keyword}

\begin{abstract}
In this paper, we study the robust consensus problem for a set of
discrete-time linear agents to coordinate over an uncertain
communication network, which is to achieve consensus against the
transmission errors and noises resulted from the information
exchange between the agents. We model the network by means of
communication links subject to multiplicative stochastic
uncertainties, which are
susceptible to describing packet dropout, random delay, and fading
phenomena.
Different communication topologies, such as undirected
graphs and leader-follower graphs, are considered.
We derive sufficient conditions for robust consensus in the mean
square sense. This results unveil {intrinsic constraints} on
consensus attainment imposed by the network synchronizability, the
unstable agent dynamics, and the channel uncertainty variances. 
Consensus protocols are designed based on the state information
transmitted over the uncertain channels, by solving a modified
algebraic Riccati equation.
\end{abstract}

\end{frontmatter}

\section{Introduction}
Over the last decade, consensus and other cooperative control
problems of multi-agent systems have
received compelling attention from the control community; 
see the recent works
\cite{RenBeard07_Springer,olfati-saber2007consensus,antonelli2013interconnected,li2014cooperative}
and the references therein. Roughly speaking, a consensus problem of
multi-agent systems is composed of three essential components, both
independent and inter-related: the agent dynamics, the network
communicating among the agents, and the consensus protocol required
to achieve consensus \cite{li2014cooperative}. The network topology
is without any question the central mechanism that enables multiple
agents to cooperate via information exchange between the agents and
over the network. In the literature, a directed or undirected graph,
commonly represented by a known, often constant matrix, is adopted
to characterize the information flow among the agents in a highly
idealized manner. In practice, however, the information exchanges
among agents must be executed over a communication network
consisting of various communication links, which cannot be immune to
transmission errors, channel noises, and system constraints. A
realistic multi-agent model must therefore account for the network
communication errors and noises, and for it to be viable, the theory
of cooperative control of multi-agent systems must address the
robustness against the noises and uncertainties in the communication
channels.

Communication errors may result from a variety of channel
constraints and transmission imperfections, such as data rate limit,
quantization precision, time delay, packet dropout, and channel
capacity, etc. These issues pose novel challenges to feedback design
and have been a focal subject of the recent research on networked
control systems
\cite{goodwin2006introduction,nair2007feedback,hespanha2007survey}.
By now it is widely known that for a system controlled over a
communication channel, the channel specification, be it data rate,
capacity, quantization density, and signal-to-noise ratio, must in
one way or another satisfy an intrinsic bound in order to ensure
feedback stabilization
\cite{tatikonda2004control,braslavsky2007feedback,nair2000stabilization}.
Unsurprisingly, these advances have made their way into the study of
multi-agent systems. Consensus problems using quantized information
exchanges are considered in, e.g.,
\cite{carli2008communication,li2010distributed}. Conditions are
derived in \cite{li2014multi,cheng2011necessary} to achieve mean
square consensus in the presence of communication or measurement
noises. Efforts are made in \cite{fagnani2009average,wu2012average}
to address the effect of data loss or packet dropout on consensus.
Generally, in the presence of such communication errors, one finds
the design of a robust consensus protocol ever more difficult and
indeed a highly nontrivial task, due to the interplay between the
agent dynamics, the communication graph, and the communication
channels.

In this paper we study the robust consensus problem for general
discrete-time multi-agent systems with stochastic uncertain
communication channels. Motivated by the random quantizers in
\cite{xiao2009mean} and the fading channel model in
\cite{elia2005remote}, we model each communication link between
the agents as a noisy communication channel subject to a
multiplicative stochastic uncertainty. A graph network so formulated
provides a more realistic model of the communication network consisting of erasure channels, which are
particularly susceptible to describing such network losses as packet
dropout and random delay. Indeed, the uncertain channel model herein
includes the random quantizers (as pointed out in
\cite{xiao2009mean}), the stochastic i.i.d. packet loss (as
demonstrated in \cite{elia2005remote}), and random
relative-state-dependent measurement noises (as shown in
\cite{li2014multi}) as special cases. Our task is twofold. First, we
design robust consensus protocols so that the agents can
achieve consensus despite the presence of the channel uncertainties.
Second and perhaps more importantly, we investigate the fundamental
limits of the uncertainties under which consensus remains
attainable. We address these issues for two different
communication topologies, including undirected graphs and
leader-follower graphs. The following passages
summarize our main contributions.

We
formulate and solve the robust consensus problem in the mean square
sense. Here we say that a group of agents achieve mean square
consensus if the states of the agents converge asymptotically to a
common state under the mean square criterion. We derive sufficient conditions 
guaranteeing the mean square consensus, under which accordingly, a
robust consensus protocol is designed based on the relative state
information; the latter is achieved by solving a modified algebraic Riccati equation. This
result sheds important light into the {intrinsic limit} to the
channel uncertainty variance that a multi-agent system can tolerate
to achieve consensus with the given graph topology and the feedback
protocol, which is seen to depend on the graph synchronizability,
the topological entropy of the agents, and the uncertainty variance.
Interestingly, the condition recovers the conventional results on
consensus of discrete-time linear multi-agent systems with ideal
channels in, e.g.,
\cite{li2011consensusDCDS,you2011network,hengster2013synchronization},
which is both necessary and sufficient in the noise-free case.

A further expansion of these results concerns the case where
uncertainties are present in the control input channels of the
agents. In this case, each agent communicates with its neighbors
through the same communication channel perturbed by a multiplicative
uncertainty. A sufficient condition is obtained similarly to guarantee the mean square consensus.

Previous works pertinent to this paper include
\cite{wang2012distributed,zelazo2014robustness,li2014multi,ma2014mean,long2015distributed},
where robust consensus over uncertain communication channels is studied. It should be noted, nonetheless, that only single-integrator agents are considered in these works, 
while the present paper
addresses general high-order linear agent dynamics. This difference
in the agent dynamics is by no means simple. Indeed, even with
single-integrator agents, the robust consensus conditions (see,
e.g., \cite{wang2012distributed}) appear rather complex, lest the
conditions for general linear agents. Equally difficult is the task
of designing feedback gain matrices of the consensus protocols. This
latter issue is moot with the aforementioned case of
single-integrator agents, as therein only scalar feedback gains need
to be designed. 
More generally, due to the need to incorporate the graph structure, the analysis and design of robust consensus against the multiplicative stochastic uncertainties presents in general a highly nontrivial task, whose difficulty goes considerably beyond that found in networked control problems \cite{goodwin2006introduction,nair2007feedback,hespanha2007survey}.
Finally, it is worth pointing out that robustness issues will also arise from the uncertainties in agent dynamics. This, however, is not the concern of the current paper; there is a bulk of research devoted to the study of heterogeneous agents, see, e.g., \cite{trentelman2013robust} and references therein.

The mathematic preliminaries used in this paper are summarized in Section \ref{s2}.
The uncertain communication channel model is presented and
the robust consensus problem is formulated in Section \ref{s3}.
The robust consensus problem in the presence of stochastic channel uncertainties
is studied in Section \ref{s4}.
The uncertain input channel model is discussed in Section \ref{s5}.
Numerical simulation results are presented for illustration in Section \ref{s6}.
Finally, Section \ref{s7} concludes our paper.

\section{Mathematical Preliminaries}\label{s2}

\subsection{Notations}

The notations used in this paper are standard. 
$\mathbf{R}^{n\times m}$ denotes the set of  $n\times m$
real matrices.
The symbol
$\mathbf{1}$ denotes a column vector with all entries equal to 1.
$\|x\|$ denotes the 2-norm of a vector $x$ and $\|A\|$ denotes the induced 2-norm of a real matrix $A$.
The matrix inequality $A>B$ means $A$ and $B$ are symmetric matrices and $A-B$ is positive definite.
$A\otimes B$ represents the Kronecker product of matrices $A$ and $B$. The expectation operator is denoted by ${\mathbf{E}\{\cdot}\}$.

\subsection{Algebraic Graph Theory}

This subsection summarizes some relevant facts on graph
theory, which are mainly adopted from
\cite{mesbahi2010graph,li2014cooperative}.

A directed graph $\mathcal{G}$ is defined by $\mathcal{G}=(\mathcal{V},\mathcal{E})$, where $\mathcal{V}=\{1,\cdots,N\}$ is the set of nodes
and $\mathcal{E}\subseteq \mathcal{V}\times\mathcal{V}$ denotes the set of edges. 
For an edge $(v_i,v_j)$, node $v_i$ is called the parent node (i.e., the head),
$v_j$ is the child node (i.e., the tail), and $v_i$ is a neighbor of $v_j$.
A graph with the property that
$(v_i,v_j)\in\mathcal {E}$ implies that $(v_j, v_i)\in\mathcal {E}$ for
any $v_i,v_j\in\mathcal {V}$, {i.e., each edge is an unordered pair of nodes}, is said to be undirected.
A path from node $i_1$ to node $i_l$ is a sequence of ordered edges in the form
of $(i_k,i_{k+1})$, $k=1,\cdots,l-1$. A directed graph contains a directed spanning tree if there exists a node called the root
which has directed paths to all other nodes in the graph.
For a directed graph $\mathcal{G}$, its adjacency matrix, denoted by $\mathcal{A}=[a_{ij}]\in \mathbf{R}^{N\times N}$, is defined such that $a_{ii}=0$, $a_{ij}$ is a positive
value if $(v_j,v_i)\in\mathcal {E}$ and $a_{ij}=0$
otherwise, where
$a_{ij}$ denotes the weight for the edge $(v_j,v_i)\in\mathcal {E}$.
For an undirected graph, assume that each of its edges has a head and a tail.
The incidence matrix associated with the undirected graph,
denoted by $\mathcal{D}=[d_{ij}]\in \mathbf{R}^{|\mathcal{V}|\times |\mathcal{E}|}$,
is defined as $d_{ij}=-1$ if $v_i$ is the tail of $(v_i,v_j)$,
{$d_{ij}=1$} if $v_i$ is the head of $(v_i,v_j)$, and
$d_{ij}=0$ otherwise.
The Laplacian matrix $\mathcal{L}=[\mathcal{L}_{ij}]\in \mathbf{R}^{N\times N}$ associated with
a graph $\mathcal{G}$ is defined as $\mathcal{L}_{ii}=\sum_{j=1}^{N}a_{ij}$ and $\mathcal{L}_{ij}=-a_{ij}$, $i\neq j$.

\begin{lem}[\cite{mesbahi2010graph}]\label{lem1}
%For an unweighted and undirected graph, $\mathcal{L}=\mathcal{D}\mathcal{D}^T$. 
For an undirected graph,
$\mathcal{L}=\mathcal{D}W\mathcal{D}^T$, where {$W$ is an} $|\mathcal{E}|\times |\mathcal{E}|$
diagonal matrix, with positive $a_{ij}$ on the diagonal.
\end{lem}

\begin{lem}[\cite{RenBeard07_Springer}]\label{lem2}
Zero is an eigenvalue of $\mathcal {L}$ with $\mathbf{1}$ as a
right eigenvector and all nonzero eigenvalues have positive real
parts. Besides, zero is a simple eigenvalue of $\mathcal {L}$ if
and only if $\mathcal {G}$ has a directed spanning tree.
\end{lem}

\section{Stochastically Perturbed Channels and Problem Formulation}\label{s3}

Consider a network consisting of $N$ discrete-time linear agents.
The dynamics of the $i$-th agent are described by
\begin{equation}\label{followers}
\begin{aligned}
&{x}_{i}(k+1)=Ax_{i}(k)+Bu_{i}(k), \quad i=1,\cdots,N,
\end{aligned}
\end{equation}
where $x_i\in\mathbf{R}^n$ is the state vector and
$u_i\in\mathbf{R}$ is the control input vector of the $i$-th agent, respectively,
and the following assumption holds.

{{\bf Assumption 1}
The pair $(A,B)$ is stabilizable.}

\begin{figure}[htbp]
\centering
\includegraphics[width=0.7\linewidth]{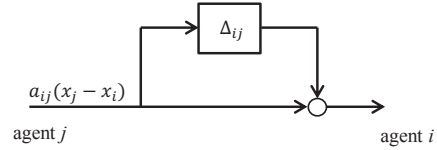}
\caption{An uncertain communication channel between agents $i$ and $j$. }\label{fig1}
\end{figure}

The information flow among the $N$ agents is depicted by a graph
$\mathcal{G}$, which can be either directed or undirected. Agent $j$
can obtain information from agent $i$, if $(i,j)$ is an edge. 
The graph
$\mathcal{G}$ merely depicts the information flow topology among the
agents. In practice, the agents cooperate via information exchange
over a network of communication channels, which in general are
subject to transmission errors resulted from communication noises
and channel constraints. Motivated by the random quantizers in
\cite{xiao2009mean} and the fading channel model in
\cite{elia2005remote}, in this paper we model each communication
channel as an ideal transmission system subject to a multiplicative
stochastic uncertainty. Specifically, each agent obtains the
relative state information from its neighbors through uncertain
channels, which are ideal channels with a unity transfer function
perturbed by an stochastic uncertainty $\Delta_{ij}$, as shown in
Fig. \ref{fig1}.

%explain the physical meanings of uncertain channels. Each channel is subject to multiplicative uncertainties,
%which can be either stochastic or deterministic.

The control input of each agent, i.e., the consensus protocol, is designed as
\begin{equation}\label{signals}
u_i(k)=\alpha\sum_{j=1}^N a_{ij}[1+\Delta_{ij}(k)]K[x_i(k)-x_j(k)],~ i=1,\cdots,N,
\end{equation}
where $\alpha\in\mathbf{R}$ is a positive scaling factor, $K\in\mathbf{R}^{1\times n}$ is the feedback gain matrix,
$a_{ij}$ denotes the $(i,j)$-th entry of the non-weighted (i.e., $a_{ij}\in\{0,1\}$)  adjacency matrix of the graph $\mathcal{G}$, and $\Delta_{ij}(k)\in\mathbf{R}$ denotes the stochastic uncertainty associated with the communication channel between agents $i$ and $j$.

We assume that the stochastic uncertainties $\Delta_{ij}(k)$ satisfy the following assumption.

{\bf Assumption 2}
$\Delta_{ij}(k)$, $(i,j)\in\mathcal{E}$ in \dref{signals} are white noise processes, independent from each other and for each $k$,
with $\mathbf{E}\{\Delta_{ij}(k)\}=0$ and $\mathbf{E}\{\Delta_{ij}^2(k)\}=\sigma_{ij}^2$.

{\bf Remark 1}~
The stochastically perturbed communication channel in this paper is
fairly general and can be used to describe several communication
losses in digital networks, such as quantization
\cite{xiao2009mean}, signal distortion
\cite{braslavsky2007feedback}, packet drop
\cite{elia2005remote}, and multiplicative (relative-state-dependent) communication noises \cite{li2014multi}. Specifically, the binary i.i.d. packet loss
and Rice fading channel \cite {elia2005remote} can be described by a
structured bounded variance uncertainty in the form of
\dref{signals}. It is also shown in \cite{xiao2009mean} that if
$\Delta_{ij}(k)$ is uniformly distributed over a certain interval
for every $i,j$, and $k$, then the uncertainty $\Delta_{ij}(k)$ can
be viewed as a random quantization error. 
%\end{remark}

The protocol \dref{signals} solves the robust consensus problem in the mean square sense,  i.e., the mean square consensus problem, 
in the presence of the uncertain channel model described as above, if
$\lim_{k\rightarrow \infty} \mathbf{E}\{\|x_i(k)-x_j(k)\|^2\}= 0$, $i,j=1,\cdots,N$. Our objective in this paper is two-fold. 
We first determine conditions such that the agents in
\dref{followers} will achieve consensus under the protocol
\dref{signals}. This in turn leads to intrinstic limits relating
the channel uncertainty specification to the network
synchronizability and the agents' unstable dynamics. Next, we design
the feedback gain matrix $K$ in \dref{signals} such that the states
of the agents reach agreement in the mean square sense.

\section{Consensus over Uncertain Communication Channels}\label{s4}

In this section, we consider the robust consensus problem with the
stochastically perturbed channel model for different network
topologies.

\subsection{Undirected Graphs}\label{s4a}

In this subsection, we consider the case where the
communication graph $\mathcal {G}$ is undirected, which satisfies
the following assumption.

{\bf Assumption 3}
The graph $\mathcal{G}$ is undirected and connected.

Let $x(k)=[x_1(k)^T,\cdots,x_N(k)]^T$. Then, it follows from \dref{followers} and \dref{signals}
that the closed-loop network dynamics can be written as
\begin{equation}\label{networks2q}
\begin{aligned}
x(k+1)&=(I_N\otimes A+\alpha\mathcal {L}\otimes BK)x(k)\\&\quad+\alpha (I_N\otimes B)\Pi(k) (I_N\otimes K)x(k),
\end{aligned}
\end{equation}
where $\Pi(k)=[\Pi_{ij}(k)]_{N\times N}\in\mathbf{R}^{N\times N}$, with $\Pi_{ij}(k)=-a_{ij}\Delta_{ij}(k)$, $i\neq j$, and $\Pi_{ii}(k)=\sum_{j=1}^Na_{ij}\Delta_{ij}(k)$,
and $\mathcal {L}$ is the symmetric Laplacian matrix associate with $\mathcal{G}$. Evidently, $\Pi(k) {\bf 1} =0.$

Let $\xi(k)=(M\otimes I_n)x(k)$, where $M\triangleq I_N-\frac{1}{N}\mathbf{1}\mathbf{1}^T$. It is easy
to see that $0$ is a simple eigenvalue of
$M$ with $\mathbf{1}$ as the
eigenvector and 1 is the other eigenvalue with
multiplicity $N-1$. Then, it follows that $\xi(k)=0$ if and only if
$x_1(k)=\cdots=x_N(k)$. Hereafter,
we will refer to $\xi(k)$ as the consensus error. In light of the facts that $\mathcal {L} M=M\mathcal {L} =\mathcal {L} $ and $\Pi(k) M=\Pi(k) $,
we can deduce from \dref{networks2q} that $\xi(k)$ satisfies the following dynamics:
\begin{equation}\label{networks2q2}
\begin{aligned}
\xi(k+1)&=(I_N\otimes A+\alpha\mathcal {L}\otimes BK)\xi(k)\\&\quad+\alpha (M\otimes B)\Pi(k) (I_N\otimes K)\xi(k),
\end{aligned}
\end{equation}
Evidently, the consensus protocol \dref{signals} solves the mean
square consensus problem for the agents in \dref{followers}, if the
system \dref{networks2q2} is mean square stable.

Before moving forwards, we introduce the following modified algebraic
Riccati equation (MARE) \cite{schenato2007foundations}:
\begin{equation}\label{ric}
P=A^TPA-(1-\breve{\delta}^2) A^TPB(B^TPB)^{-1}B^TPA+Q,
\end{equation}
where $P\geq0$, $Q>0$, and $\breve{\delta}\in\mathbf{R}$. The lemma given below concerns the existence of a solution to the
MARE \dref{ric}.

\begin{lem}[\cite{schenato2007foundations}]\label{lric}
Assume that $(A,B)$ be stablizable and $B\in\mathbf{R}^n$.
Then, the MARE \dref{ric} has a unique positive-definite solution $P$,
if $|\breve{\delta}|<1/M(A)$, where $M(A)\triangleq\prod_{i=1}^n \max\{1,|\lambda_i(A)|\}$
denotes the Mahler measure of $A$, which is the absolute value of the product of the unstable
eigenvalues of $A$. %The converse is true if $A$ is unstable.
\end{lem}

The following theorem presents a sufficient condition to solve the mean square consensus problem.

\begin{thm}\label{thm1}
Suppose that Assumptions 1--3 hold. There exists a consensus protocol in the form of \dref{signals}
solving the mean square consensus problem for the agents in \dref{followers},
if the following condition holds:
\begin{equation}\label{concons1}
(\alpha\lambda_i-1)^2+{\alpha^2\sigma_{\max}^2\lambda_i}<\frac{1}{M(A)^2},~ i=2,N,
\end{equation}
where $\sigma_{\max}^2\triangleq\max_{(i,j)\in\mathcal{E}}\{\sigma_{ij}^2+\sigma_{ji}^2\}$
% is the largest variance of the uncertainties in all channels 
and
$\lambda_2\leq\cdots\leq\lambda_N$ are the nonzero eigenvalues of
$\mathcal {L}$. Furthermore, the feedback gain matrix $K$ in
\dref{signals} can be chosen as $K=-(B^TPB)^{-1}B^TPA$, where $P>0$
is the unique solution to the MARE \dref{ric} with
$\max_{i=2,N}\{(\alpha\lambda_i-1)^2+\alpha^2\sigma_{\max}^2\lambda_i\}\leq
\breve{\delta}^2<1/M(A)^2$.
\end{thm}

\vspace*{-5pt}
{\bf Proof}~
Because Assumption 3 holds, it follows from Lemma  \ref{lem2} that zero is a simple eigenvalue of $\mathcal {L}$ and
all the other eigenvalues are positive. Let $U\in\mathbf{R}^{N\times
N}$ be a unitary matrix such that $U^{T}\mathcal
{L}U=\Lambda\triangleq{\rm{diag}}(0,\lambda_2,\cdots,\lambda_N)$.
Since the right and left eigenvectors of $\mathcal {L}$
corresponding to the zero eigenvalue are ${\bf 1}$ and ${\bf 1}^T$,
respectively, we select
%\begin{equation}\label{unitary}
$U=\left[\begin{smallmatrix}
\frac{\mathbf{1}}{\sqrt{N}} & Y_1
\end{smallmatrix}\right]$,
%\end{equation}
where $Y_1\in\mathbf{R}^{N\times(N-1)}$.
It is not difficult to check that
\begin{equation}\label{unitary1}
\begin{aligned}
Y_1^TY_1 &=I_{N-1},~Y_1Y_1^T=M,\\
Y_1^T\mathcal {L} Y_1 &=\Lambda_1\triangleq{\rm{diag}}(\lambda_2,\cdots,\lambda_N).
\end{aligned}
\end{equation}
Let
$\tilde{\xi}(k)\triangleq[\tilde{\xi}_1(k)^T,\cdots,\tilde{\xi}_N^T(k)]^T=(U^T\otimes I_n)\xi(k)$. Then,
\dref{networks2q2} can be rewritten into
\begin{equation}\label{networks2q3}
\begin{aligned}
\tilde{\xi}(k+1)&=(I_N\otimes A+\alpha\Lambda\otimes BK)\tilde{\xi}(k)\\&\quad+\alpha (U^TM\otimes B)\Pi(k) (U\otimes K)\tilde{\xi}(k).
\end{aligned}
\end{equation}
In virtue of the definitions of $\xi(k)$ and $\tilde{\xi}(k)$, we
can verify that $
\tilde{\xi}_1(k)=(\frac{\mathbf{1}^T}{\sqrt{N}}\otimes
I_n)\xi(k)\equiv0.$ Therefore, the mean square consensus problem is
reduced to checking the mean square stability of $\zeta(k)\triangleq[\tilde{\xi}_2(k)^T,\cdots,\tilde{\xi}_N^T(k)]^T$.
%$\zeta(k)\triangleq[\tilde{\xi}_2^T(k),\cdots,\tilde{\xi}_N^T(k)]^T$.
Since $U^TM=\left[\begin{smallmatrix} 0 \\
Y_1^TM\end{smallmatrix}\right]$ and $\Pi(k)U=\left[\begin{smallmatrix} 0 &
\Pi(k)Y_1\end{smallmatrix}\right]$, it follows that
$$\begin{aligned}
(U^TM&\otimes B)\Pi(k) (U\otimes K)\\&=
{\mathrm{diag}}(0, (I\otimes B)(Y_1^T\Pi(k)Y_1)(I\otimes K)),
\end{aligned}$$
where we have used the fact that $Y_1^TM=Y_1$, which follows readily from \dref{unitary1}.
Then, we can derive from \dref{networks2q3} that 
$\zeta(k)$ satisfies
\begin{equation}\label{networks3u}
\begin{aligned}
\zeta(k+1) =&(I\otimes A+\alpha\Lambda_1\otimes BK)\zeta(k)+\alpha (I\otimes B)\\&\times(Y_1^T\Pi(k)Y_1)(I\otimes K)\zeta(k).
\end{aligned}
\end{equation}

By using a slightly modified version of Theorem 6.4 in \cite{elia2005remote} or the result in Section 9.1.1 in \cite{boyd1994linear},  the system \dref{networks3u} in mean square stable if and only if there exists a $\mathcal{P}>0$ such that
\begin{equation}\label{th2equ0}
\begin{aligned}
&\mathcal{P}-(I\otimes A^T+\alpha\Lambda_1\otimes K^TB^T)\mathcal{P}(I\otimes A-\alpha\Lambda_1\otimes BK)\\&\quad-\alpha^2(I\otimes K^T)\mathbf{E}\{Y_1^T\Pi(k)^TY_1(I\otimes B^T)\mathcal{P}\\&\quad\times (I\otimes B)Y_1^T\Pi(k)Y_1\} (I\otimes K)>0. 
\end{aligned}
\end{equation}
In  the following, we will show that $I\otimes P$, with $P>0$ being the solution to the MARE \dref{ric}, satisfies \dref{th2equ0}. With $\mathcal{P}=I\otimes P$, the inequality \dref{th2equ0} can be rewritten as 
\begin{equation}\label{th2equ0x}
\begin{aligned}
&I\otimes P-(I\otimes A^T+\alpha\Lambda_1\otimes K^TB^T)(I\otimes P)(I\otimes A\\
&\quad+\alpha\Lambda_1\otimes BK)-\alpha^2\mathbf{E}\{Y_1^T\Pi(k)^TY_1Y_1^T\Pi(k)Y_1\} \\&\quad\otimes K^TB^TPBK>0. 
\end{aligned}
\end{equation}
By noting that $M\leq I$ and in light of \dref{unitary1}, we have 
\begin{equation}\label{th2equ2}
\begin{aligned}
Y_1^T\Pi(k)^TY_1Y_1^T\Pi(k)Y_1&=Y_1^T\Pi(k)^TM\Pi(k)Y_1\\&\leq Y_1^T\Pi(k)^T\Pi(k)Y_1.
\end{aligned}
\end{equation}
Under Assumption 2, it is not difficult to derive that the expectation of $\Pi(k)^T\Pi(k)$ satisfies 
\begin{equation}\label{th2equ4}
\begin{aligned}
&\mathbf{E}\{\Pi(k)^T\Pi(k)\}=\Theta,\\
&\theta_{ij}={-}a_{ij}(\sigma_{ij}^2+\sigma_{ji}^2), ~i\neq j,\\
&\theta_{ii}=\sum_{j=1}^Na_{ij}(\sigma_{ji}^2 +\sigma_{ij}^2),~i=1,\cdots,N,
\end{aligned}
\end{equation}
where $\theta_{ij}$ denotes the $(i,j)$-th entry of the matrix $\Theta$ and we have also used the fact that $a_{ij}=a_{ji}$.
It is not difficult to verify that $\Theta$ corresponds to certain weighted Laplacian matrix of the undirected graph $\mathcal{G}$.
Suppose that there are $M$ edges in the graph $\mathcal {G}$
and denote by $\mathcal {D}\in\mathbf{R}^{N\times M}$ the incidence matrix of $\mathcal {G}$.
In light of Lemma 1, we can verify that 
\begin{equation}\label{th2equ7}
\begin{aligned}
\Theta&=\mathcal{D}\Delta \mathcal{D}^T,\\
\Delta &={\rm{diag}}(\sigma_{ji}^2 +\sigma_{ij}^2,\forall(i,j)\in\mathcal {E})\in\mathbf{R}^{M\times M}.
\end{aligned}
\end{equation}
Then, we can derive from \dref{th2equ2},  \dref{th2equ4}, and \dref{th2equ7} that
\begin{equation}\label{th2equ7x}
\begin{aligned}
\mathbf{E}\{Y_1^T\Pi(k)^TY_1Y_1^T\Pi(k)Y_1\}&\leq {\sigma_{\max}^2 Y_1^T\mathcal{D}\mathcal{D}^TY_1}\\&=\sigma_{\max}^2 Y_1^T\mathcal{L}Y_1\\& {=\sigma_{\max}^2\Lambda_1}.
\end{aligned}
\end{equation}
By invoking \dref{th2equ7x}, it follows that \dref{th2equ0x} holds, if $\mathcal{M} >0$,
where 
\begin{equation}\label{th2equ1x2}
\begin{aligned}
\mathcal{M}=&I\otimes P-(I\otimes A^T+\alpha\Lambda_1\otimes K^TB^T)(I\otimes P)(I\otimes A\\&+\alpha\Lambda_1\otimes BK)
-{\alpha^2\sigma_{\max}^2 \Lambda_1}\otimes K^TB^TPBK.
\end{aligned}
\end{equation}
Since the matrices on the right side of \dref{th2equ1x2} are block diagonal, it is evident that 
$\mathcal{M}>0$, if 
\begin{equation}\label{th2equ7y}
\begin{aligned}
P&>(A+\alpha\lambda_iBK)^TP(A+\alpha\lambda_iBK)\\&\quad+\alpha^2\sigma_{\max}^2\lambda_i K^TB^TPBK, ~ i=2,\cdots,N.
\end{aligned}
\end{equation}
Substituting $K=-(B^TPB)^{-1}B^TPA$ into \dref{th2equ7y} and invoking
\dref{concons1} yield
\begin{equation}\label{th2equ8}
\begin{aligned}
&P-A^TPA-[(\alpha\lambda_i-1)^2-1+\sigma_{\max}^2\alpha^2\lambda_i]A^TPB\\&\quad\times(B^TPB)^{-1}B^TPA\\
&\geq P-A^TPA+(1-\breve{\delta}^2) A^TPB(B^TPB)^{-1}B^TPA>0, %\quad i=1,\cdots,N.
\end{aligned}
\end{equation}
where we have used the fact that
$(\alpha\lambda_i-1)^2-1+\sigma_{\max}^2\alpha^2\lambda_i$ is
convex with respect to $\lambda_i$. In light of \dref{th2equ7y} and \dref{th2equ8}, it follows from \dref{th2equ1x2} that 
%\begin{equation}\label{th2equ7y}
%\begin{aligned}
$\mathcal{M}>0.$
%\end{aligned}
%\end{equation}
Then, we obtain that $\mathcal{P}=I\otimes P$ satisfies \dref{th2equ0}, 
i.e., the system
\dref{networks3u} is mean square stable. Therefore,
the consensus protocol \dref{signals} with $K$ chosen as above
solves the mean square consensus problem provided that the condition
\dref{concons1} holds. \hfill $\blacksquare$

{\bf Remark 2}~
The scalar $\alpha$ in \dref{signals} is a design scaling parameter,
which is intended to regulate the left-hand side of \dref{concons1}.
{By simple calculations, we can
choose $\alpha=\frac{2}{\lambda_2+\lambda_N+\sigma_{\max}^2}$ such
that left-hand side of \dref{concons1} is minimized for given $\lambda_2$, $\lambda_N$,
and $\sigma_{\max}^2$.  In this case, the condition
\dref{concons1} becomes
\begin{equation}\label{concons1n}
%\frac{(1-\frac{\lambda_2}{\lambda_N}-\frac{\sigma_{\max}^2}{\lambda_N})^2}{(1+\frac{\lambda_2}{\lambda_N}+\frac{\sigma_{\max}^2}{\lambda_N})^2}+\frac{4\frac{\sigma_{\max}^2}{\lambda_N}}{(1+\frac{\lambda_2}{\lambda_N}+\frac{\sigma_{\max}^2}{\lambda_N})^2}<\frac{1}{M(A)^2}.
%\frac{(1-\frac{\lambda_2}{\lambda_N}-\frac{\sigma_{\max}^2}{\lambda_N})^2+4\frac{\sigma_{\max}^2}{\lambda_N}}{(1+\frac{\lambda_2}{\lambda_N}+\frac{\sigma_{\max}^2}{\lambda_N})^2}<\frac{1}{M(A)^2}.
\frac{(\lambda_N- \lambda_2-\sigma_{\max}^2)^2+4\sigma_{\max}^2\lambda_N}{(\lambda_N+ \lambda_2+\sigma_{\max}^2)^2}<\frac{1}{M(A)^2}.
\end{equation}}
On the other hand, If the variances of the channel uncertainties are not available, in this case we can choose
 $\alpha=\frac{2}{\lambda_2+\lambda_N}$ and  the condition
\dref{concons1} becomes
 \begin{equation}\label{concons1n2}
\frac{(\lambda_N- \lambda_2)^2+4\sigma_{\max}^2\lambda_N}{(\lambda_N+ \lambda_2)^2}<\frac{1}{M(A)^2}.
\end{equation}
When the communication channels are noise-free, which is equivalent to   $\sigma_{ij}^2=0$ for all $i,j$, 
the condition \dref{concons1n} or \dref{concons1n2}  reduces to
$\frac{1-\lambda_2/\lambda_N}{1+\lambda_2/\lambda_N}<\frac{1}{M(A)}$. The latter
 is the necessary and sufficient condition for consensus with
ideal channels, obtained in, e.g.,
\cite{li2011consensusDCDS,you2011network,hengster2013synchronization}.

{\bf Remark 3}~
%The fundamental limitation caused the uncertain channels.
Theorem \ref{thm1} uncovers the intrinsic constraints induced by the channel uncertainties, the communication topology, and
the unstable dynamics of the agents. The eigenratio $\lambda_2/\lambda_N$ in \dref{concons1n}
is commonly used to indicate the synchronizability of dynamical networks \cite{duan2007complex,pecora1998master}. A
larger eigenratio implies a better synchronizability. 
The Mahler measure $M(A)$ of $A$, or the topological entropy $\log_2
M(A)$, plays an important role in networked control systems
\cite{nair2007feedback,qiu2013stabilization}, as it describes the
minimum data rate for stabilizing an unstable system. The largest
variance $\sigma_{\max}^2$ can be used to measure the maximum mean
square channel capacity for the stochastic communication channels,
as discussed in \cite{elia2005remote}. As demonstrated in the
condition \dref{concons1n}, the synchronizability factor, the
topological entropy of $A$, and the mean square channel capacity
impose { intrinsic limitations} on achieving consensus over
stochastic uncertain channels.

{{\bf Remark 4}~
For the case where the agents have no eigenvalues outside the unit circle, i.e., $M(A)=1$,
it is worth noting that the condition \dref{concons1n} always holds and the consensus protocol \dref{signals}  can be constructed as in Theorem 1 and Remark 2 to ensure mean square consensus for any channel uncertainties satisfying Assumption 3 as long as $(A,B)$ is stabilizable. In this sense,
the condition \dref{concons1n} is both necessary and sufficient. Nevertheless,
the conservatism of the condition  \dref{concons1n} remains unclear for the case of $M(A)>1$.
The condition \dref{concons1n2} is generally not necessary.
For instance, consider a group of single integrators with a complete graph and assume for simplicity that the variances of the channel uncertainties are all equal, denoted by $\bar{\sigma}^2$. In this case, the condition \dref{concons1n2} becomes 
$\bar{\sigma}^2<\frac{N}{2}$.
In light of the frequency-domain mean-square small gain theorem in \cite{elia2005remote,lu2002mean},  
a necessary and sufficient condition to achieve mean square consensus is that $\bar{\sigma}^2<1$ when $N=2$,
$\bar{\sigma}^2<1.5$ when $N=3$, and $\bar{\sigma}^2<16/7$ when $N=4$.
It is easy to verify that the condition \dref{concons1n2} is tight 
for the special cases with two or three single-integrator agents with complete graphs,
and is conservative if $N\geq 4$. }

\subsection{Leader-follower Graphs}

In this subsection, we consider the case where there exists a
leader. Without loss of generality, let the agent in
\dref{followers} indexed by $1$ be the leader and the rest be the
followers. The leader receives no information from any follower and
only a subset of the followers have access to the leader's
information. It is assumed that the leader's control input is equal
to zero, i.e., $u_1=0$, and the control inputs of the followers are
described by \dref{signals}. We assume that the communication graph
$\mathcal {G}$ among the $N$ agents satisfies the following
assumption.

{\bf Assumption 4}  
The graph $\mathcal{G}$ contains a directed spanning tree with the leader as the root
and the subgraph $\mathcal{G}_s$ associated with
the {$N-1$} followers is undirected.

Because the leader has no neighbors, the Laplacian matrix $\mathcal {L}$
associated with
$\mathcal {G}$ can be partitioned as
%\begin{equation}\label{lapc}
$\mathcal {L}=\left[\begin{smallmatrix} 0 & 0_{1\times (N-1)} \\
\mathcal {L}_2 & \mathcal {L}_1\end{smallmatrix}\right]$,
%\end{equation}
where $\mathcal {L}_2\in\mathbf{R}^{(N-1)\times 1}$ and $\mathcal
{L}_1\in\mathbf{R}^{(N-1)\times (N-1)}$.
By Lemma \ref{lem2} and Assumption 4, it is clear that
$\mathcal {L}_1>0$.

Let $e_i(k)=x_i(k)-x_1(k)$, $i=2,\cdots,N$. From \dref{followers}
and \dref{signals}, we find that
\begin{equation}\label{networks}
e_i(k+1)=Ae_i(k)+\alpha B\sum_{j=1}^N a_{ij}[1+\Delta_{ij}(k)]K[e_i(k)-e_j(k)],%\quad i=1,\cdots,N,
\end{equation}
where $e_1=0$. Let $e(k)=[e_2(k)^T,\cdots,e_N(k)]^T$. Then, we rewrite \dref{networks} into
\begin{equation}\label{networks2}
\begin{aligned}
e(k+1)&=(I_{N-1}\otimes A+\alpha\mathcal {L}_1\otimes BK)e(k)\\&\quad+\alpha (I_{N-1}\otimes B)\widehat{\Pi}(k) (I_{N-1}\otimes K)e(k),
\end{aligned}
\end{equation}
where $\widehat{\Pi}(k)=[\widehat{\Pi}_{ij}(k)]_{(N-1)\times
(N-1)}$, with $\widehat{\Pi}_{ij}(k)=-a_{ij}\Delta_{ij}(k)$, $i\neq
j$, and $\widehat{\Pi}_{ii}(k)=\sum_{j=1}^Na_{ij}\Delta_{ij}(k)$.
Evidently, $x_i(k)$, $i=2,\cdots,N$, approach
$x_1(k)$, if $e(k)$ in \dref{networks2} converges to
zero. Therefore, the consensus protocol \dref{signals} solves the
mean square leader-follower consensus problem for the agents in
\dref{followers}, if the system \dref{networks2} is mean square
stable.

\begin{thm}\label{thm1lf}
Suppose that Assumptions 1, 2, and 4 hold. 
%Let {$\alpha=\frac{2}{\lambda_2+\lambda_N+\tilde{\sigma}_{\max}^2}$, where $\tilde{\sigma}_{\max}^2=\underset{i,j=2,\cdots,N}{\max}\{\sigma_{j1}^2,  \sigma_{ij}^2+\sigma_{ji}^2\}$}. 
Then
there exists a consensus protocol \dref{signals} achieving mean square
consensus for the agents in \dref{followers}, if 
{\begin{equation}\label{concons1nlf}
%\frac{(1-\lambda_2/\lambda_N)^2}{(1+\lambda_2/\lambda_N)^2}+\frac{8\tilde{\sigma}_{\max}^2}{\lambda_N(1+\lambda_2/\lambda_N)^2}<\frac{1}{M(A)^2}.
%\frac{(1-\frac{\lambda_2}{\lambda_N}-\frac{\tilde{\sigma}_{\max}^2}{\lambda_N})^2}{(1+\frac{\lambda_2}{\lambda_N}+\frac{\tilde{\sigma}_{\max}^2}{\lambda_N})^2}+\frac{4\frac{\sigma_{\max}^2}{\lambda_N}}{(1+\frac{\lambda_2}{\lambda_N}+\frac{\tilde{\sigma}_{\max}^2}{\lambda_N})^2}<\frac{1}{M(A)^2}.
%\frac{(\lambda_N- \lambda_2-\tilde{\sigma}_{\max}^2)^2+4\sigma_{\max}^2\lambda_N}{(\lambda_N+ \lambda_2+\tilde{\sigma}_{\max}^2)^2}<\frac{1}{M(A)^2}
(\alpha\lambda_i-1)^2+\alpha^2\tilde{\sigma}_{\max}^2\lambda_i<\frac{1}{M(A)^2},~ i=2,N,
\end{equation}}
where {$\tilde{\sigma}_{\max}^2\triangleq\underset{i,j=2,\cdots,N}{\max}\{\sigma_{j1}^2,  \sigma_{ij}^2+\sigma_{ji}^2\}$}.
\end{thm}
\vspace*{-5pt}
{\bf Proof}~Similarly as in the proof of Theorem \ref{thm1},
the system \dref{networks2} is mean square
stable, if the following inequality holds:
\begin{equation}\label{th3equ1x}
\begin{aligned}
&(I\otimes A^T+\alpha\mathcal {L}_1\otimes K^TB^T)(I\otimes P)(I\otimes A+\alpha\mathcal {L}_1\otimes BK)\\&\quad-(I\otimes P)+\alpha^2 \mathbf{E}\{\widehat{\Pi}(k)^T\widehat{\Pi}(k)\}\otimes K^TB^TPBK<0,
\end{aligned}
\end{equation}
where $K=-(B^TPB)^{-1}B^TPA$ and $P>0$ is a solution to the MARE \dref{ric} with 
$\breve{\delta}^2$ lies within the left-hand and right-hand sides of \dref{concons1nlf}.
%$\frac{(1-\lambda_2/\lambda_N)^2}{(1+\lambda_2/\lambda_N)^2}+\frac{8\tilde{\sigma}_{\max}^2}{\lambda_N(1+\lambda_2/\lambda_N)^2}\leq
%\breve{\delta}^2<\frac{1}{M(A)^2}$.
Under Assumption 2, it is not difficult to derive that 
\begin{equation}\label{th3equ4x}
\begin{aligned}
& \mathbf{E}\{\widehat{\Pi}(k)^T\widehat{\Pi}(k)\}=\widehat{\Theta}\in\mathbf{R}^{(N-1)\times (N-1)},\\
&\hat{\theta}_{ij}={-}a_{ij}(\sigma_{ij}^2+\sigma_{ji}^2), ~i\neq j,\\
&\hat{\theta}_{ii}=\sum_{j=2}^N[a_{ij}(\sigma_{ji}^2 +\sigma_{ij}^2)+a_{j1}\sigma_{j1}^2],~i=2,\cdots,N,
\end{aligned}
\end{equation}
where $\hat{\theta}_{ij}$ denotes the $(i,j)$-th entry of the matrix $\widehat{\Theta}$.

Without loss of generality, assume that the followers indexed by $2,\cdots,q+1$, have access to the leader.
Further, suppose that there are $M$  edges
in the graph $\mathcal {G}$. Thus, there exist $M-q$ edges in the subgraph $\mathcal {G}_s$
among the $N-1$ followers. Let
%\begin{equation}\label{uncersd}
$\widehat{\Delta}=\mathrm{diag}(\underbrace{\sigma^2_{21},\cdots, \sigma^2_{(q+1)1}}_{q}, \underbrace{\sigma_{ji}^2 +\sigma_{ij}^2,\forall(i,j)\in\mathcal{G}_s}_{M-q}),$
%\end{equation}
where the last $M-q$ items correspond to the $M-q$ channels in $\mathcal {G}_s$. Define $\overline{\mathcal {D}}=\begin{bmatrix} \bar{B} & \mathcal {D}_s \end{bmatrix}$,
%\begin{equation}\label{incide}
%\overline{\mathcal {D}}=\begin{bmatrix} \bar{B} & \mathcal {D}_s \end{bmatrix},
%\end{equation}
where $\mathcal {D}_s\in\mathbf{R}^{N\times (M-q)}$ denotes the incidence matrix associated with $\mathcal {G}_s$
and $\bar{B}\in\mathbf{R}^{N\times q}$ is formed by the first $q$ columns of $I_N$.

We first show the following claim.

{\bf Claim 1}~
The matrix $\widehat{\Theta}$, defined as in \dref{th3equ4x}, is equal to $\overline{\mathcal {D}}\widehat{\Delta}\overline{\mathcal {D}}^T$.

Let $\widehat{\Delta}_1=\mathrm{diag}(\sigma^2_{21},\cdots, \sigma^2_{(q+1)1})\in\mathbf{R}^{q\times q}$ and
$\widehat{\Delta}_2=\mathrm{diag}(\sigma_{ji}^2 +\sigma_{ij}^2,\forall(i,j)\in\mathcal{G}_s)\in\mathbf{R}^{(M-q)\times (M-q)}$. Denote by
$\mathcal {L}_s$ the weighted Laplacian matrix corresponding to the undirected subgraph $\mathcal {G}_s$, which,
in virtue of Lemma \ref{lem1}, is equal to $\mathcal {D}_s\widehat{\Delta}_2\mathcal {D}_s^T$. Then,
\begin{equation}\label{esxg}
\begin{aligned}
\overline{\mathcal {D}}\widehat{\Delta} \overline{\mathcal {D}}^T &=\begin{bmatrix} \bar{B} & \mathcal {D}_s \end{bmatrix}
\begin{bmatrix} \widehat{\Delta}_1  & 0\\0 &\widehat{\Delta}_2  \end{bmatrix}\begin{bmatrix} \bar{B}^T \\ \mathcal {D}_s^T \end{bmatrix}\\
&=\bar{B}\widehat{\Delta}_1\bar{B}^T+\mathcal {D}_s\widehat{\Delta}_2\mathcal {D}_s^T\\
&=\mathrm{diag}(\sigma^2_{21},\cdots, \sigma^2_{(q+1)1},0,\cdots,0)+\mathcal {L}_s.
\end{aligned}
\end{equation}
By the definition of $\widehat{\Theta}$, it is not difficult to
verify that the right hand of the last equation in \dref{esxg} is
equal to $\widehat{\Theta}$.

In light of Claim 1, it follows from \dref{th3equ4x} that 
 \begin{equation}\label{th3equ7x}
\begin{aligned}
\mathbf{E}\{\widehat{\Pi}(k)^T\widehat{\Pi}(k)\}&=\overline{\mathcal {D}}\widehat{\Delta}\overline{\mathcal {D}}^T\leq \tilde{\sigma}_{\max}^2 \overline{\mathcal {D}}\overline{\mathcal {D}}^T\\&={ \tilde{\sigma}_{\max}^2 \mathcal{L}_1}.
\end{aligned}
\end{equation}
Note that $\overline{\mathcal {D}}\overline{\mathcal {D}}^T=\mathcal {L}_1$ is a direct consequence of Claim 1 with $\widehat{\Delta}=I$.
In light of \dref{th3equ7x}, we know that \dref{th3equ1x} holds, if 
\begin{equation}\label{th3equ1y}
\begin{aligned}
&(I\otimes A^T+\alpha\mathcal {L}_1\otimes K^TB^T)(I\otimes P)(I\otimes A+\alpha\mathcal {L}_1\otimes BK)\\&\quad-(I\otimes P)+\alpha^2  \tilde{\sigma}_{\max}^2 \mathcal{L}_1\otimes K^TB^TPBK<0.
\end{aligned}
\end{equation}
Following similar steps in the proof of Theorem \ref{thm1}, it is not difficult to show that if the condition \dref{concons1nlf} holds, then
\dref{th3equ1y} and also \dref{th3equ1x} hold, implying that  \dref{networks2} is mean square
stable, i.e., 
the consensus protocol \dref{signals} solves the
mean square leader-follower consensus problem. The details are omitted here for conciseness. 
\hfill $\blacksquare$

{\bf Remark 5}~For the leader-follower graphs considered in this subsection,  communications among the followers are bidirectional, while those between the followers and the leader are directional. Therefore, it is reasonable to define the maximum mean square channel capacity in this case by the maximum of the sum of the variances of the stochastic uncertainties on every bidirectional channel and the variances of the uncertainties on the directional channels. This is different from the undirected graph case in the previous subsection.

%{\bf Remark 4}~For the leader-follower graphs considered in this subsection, the maximum mean square channel capacity is defined to be the maximum of the variances of the stochastic uncertainties on the bidirectional channels and half of the variances of the uncertainties on the directional channels. This is different from the undirected graph case in the previous subsection.

\section{Consensus via Uncertain Input Channels}\label{s5}

\begin{figure}[htbp]
\centering
\includegraphics[width=0.5\linewidth]{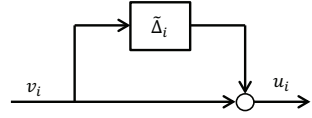}
\caption{The uncertain input channel for agent $i$. }\label{fig4}
\end{figure}

In this section, we consider the case where the agents in \dref{followers} are perturbed by uncertainties in their control input channels.
The stablization problem of a single LTI system with uncertain input channels has been considered in \cite{qiu2013stabilization,xiao2012feedback}.
The uncertain input channel model is illustrated in Fig. \ref{fig4}, where
$\tilde{\Delta}_i(k)$ denotes the stochastic uncertainty in the input channel of agent $i$ and
$v_i(k)=\sum_{j=1}^N a_{ij}K[x_i(k)-x_j(k)]$ denotes the signal available to agent $i$, which will go through the uncertain input channel for controller design. The stochastic uncertainties $\tilde{\Delta}_{i}(k)$ satisfy the following assumption.

{\bf Assumption 5}  
$\tilde{\Delta}_{i}(k)$, $i=1,\cdots,N$, in \dref{signali} are independent zero-mean white noise processes
with variances $\tilde{\sigma}_{i}^2$.

In this section, for simplicity we only consider the case where the communication graph $\mathcal {G}$
is undirected and satisfies Assumption 3.
The consensus protocol is given as follows:
\begin{equation}\label{signali}
u_i(k)=\alpha\sum_{j=1}^N a_{ij}[1+\tilde{\Delta}_{i}(k)]K[x_i(k)-x_j(k)],i=1,\cdots,N,
\end{equation}
where $a_{ij}$, $\alpha$, and $K$ are defined as in \dref{signals}.

This uncertain input channel model is related to the uncertain communication channel in the previous sections. By comparing \dref{signali} to \dref{signals}, we can observe that the uncertain input channel model can be regarded in the sense that each agent communicates with its neighbors through the same communication channel perturbed by a multiplicative uncertainty $1+\tilde{\Delta}_i(k)$.
The uncertain channels are different for different agents, even though the agents may have bidirectional communication links. This differs from the uncertain communication channel model in the previous sections.

Let $x(k)=[x_1(k)^T,\cdots,x_N(k)^T]^T$. It then follows from \dref{followers}
and \dref{signali} that $x(k)$ in this case satisfies
\begin{equation}\label{networks2gi}
x(k+1)=[I\otimes A+\alpha\mathcal {L}\otimes BK+\alpha (I \otimes B)\widetilde{\Delta}(k) (\mathcal {L}\otimes K)]x(k),
\end{equation}
where $\widetilde{\Delta}(k)\triangleq \mathrm{diag}(\tilde{\Delta}_1(k),\cdots,\tilde{\Delta}_N(k))$ and $\mathcal {L}$ is the Laplacian matrix of
$\mathcal {G}$.

The following theorem ensures that \dref{signali} solves the mean square consensus problem.

\begin{thm}\label{thm21}
Suppose that Assumptions 1, 2, and 5 hold. 
%Let {$\alpha=\frac{2}{\lambda_2+\lambda_N+\varrho^2}$, where $\varrho^2\triangleq\max\,\tilde{\sigma}_{i}^2$}.
The consensus protocol \dref{signali}
solves the mean square consensus problem for the agents in \dref{followers}, if
%\begin{equation}\label{concons1nx}
%$$\frac{(1-\lambda_2/\lambda_N-\varrho^2/\lambda_N)^2}{(1+\lambda_2/\lambda_N+\varrho^2/\lambda_N)^2}+\frac{4\varrho^2/\lambda_N}{(1+\lambda_2/\lambda_N+\varrho^2/\lambda_N)^2}<\frac{1}{M(A)^2}.$$
%$$\frac{(\lambda_N-\lambda_2-\varrho^2)^2+4\varrho^2\lambda_N}{(\lambda_N+\lambda_2+\varrho^2)^2}<\frac{1}{M(A)^2}.$$
%\end{equation}
$$(\alpha\lambda_i-1)^2+\alpha^2\varrho^2\lambda_i<\frac{1}{M(A)^2},~ i=2,N,$$
where $\varrho^2\triangleq\max\,\tilde{\sigma}_{i}^2$.
\end{thm}

\vspace*{-5pt}
{\bf Proof}~
Let the consensus error $\xi(k)$ be defined as in \dref{networks2q2}. We can verify that $\xi(k)$ in this case satisfies
\begin{equation}\label{networks2q2c}
\begin{aligned}
\xi(k+1)&=(I\otimes A+\alpha\mathcal {L}\otimes BK)\xi(k)\\&\quad+\alpha (M\otimes B)\Delta(k)\mathcal {L} (I\otimes K)\xi(k),
\end{aligned}
\end{equation}
Since $\mathcal {L}M=M\mathcal {L}$ and
both $M$ and $\mathcal {L}$ have zero row and column sums,
we can choose $U_2\triangleq\left[\begin{smallmatrix}
\frac{\mathbf{1}}{\sqrt{N}} & Y_2
\end{smallmatrix}\right]\in\mathbf{R}^{N\times N}$, with $Y_2\in\mathbf{R}^{N\times(N-1)}$,
be the unitary matrix such that $U_2^{T}M
U_2=\widehat{M}\triangleq{\rm{diag}}(0,I_{N-1})$ and
$U_2^{T}\mathcal {L}U_2=\Lambda\triangleq{\rm{diag}}(0,\Lambda_1)$,
where $\Lambda_1={\rm{diag}}(\lambda_2,\cdots,\lambda_N)$. Let
$\check{\xi}(k)\triangleq[\check{\xi}_1(k)^T,\cdots,\check{\xi}_N^T(k)]^T=(U_2^T\otimes
I_n)\xi(k)$, where $\check{\xi}_1(k)=0$. 
Since $\widehat{M}U_2^T\Delta(k)U_2\Lambda=\left[\begin{smallmatrix}
0 & 0\\ 0 & Y_2^T\Delta(k)Y_2\Lambda_1\end{smallmatrix}\right],$ we can derive that $\check{\zeta}(k)\triangleq[\check{\xi}_2(k)^T,\cdots,\check{\xi}_N^T(k)]^T$satisfies  
\begin{equation}\label{networks2q3c}
\begin{aligned}
\check{\zeta}(k+1)
&=(I \otimes A+\alpha\Lambda\otimes BK)\check{\zeta}(k)\\&\quad+\alpha (I\otimes B)Y_2^T\Delta(k)Y_2\Lambda_1(I\otimes K)\check{\zeta}(k).
\end{aligned}
\end{equation}
The rest of the derivations can be completed by following steps in the proof of Theorems \ref{thm1}, by further noting that 
$$\begin{aligned}
\mathbf{E}\{\Lambda_1Y_2^T\Delta(k)Y_2Y_2^T\Delta(k)Y_2\Lambda_1\}&\leq \mathbf{E}\{\Lambda_1Y_2^T\Delta(k)^2Y_2\Lambda_1\}\\
&\leq \varrho^2 \Lambda_1^2,
\end{aligned}$$
where we {have} used the facts that $Y_2Y_2^T=M$ and $Y_2^TY_2=I$.
The details are omitted here for brevity. \hfill $\blacksquare$

\begin{figure}[htbp]
\centering
\includegraphics[width=0.8\linewidth,height=0.5\linewidth]{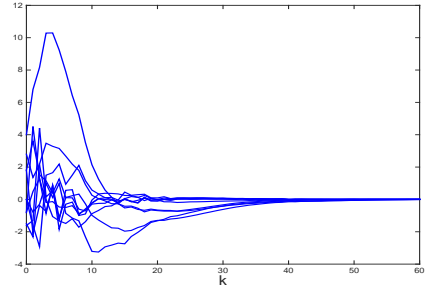}
\caption{{The average of the sampled trajectories of $x_i(k)-x_1(k)$, $i=2,\cdots,6$}.}
\label{fig5}
\end{figure}

\section{Numerical Simulations}\label{s6}

Consider a group of six discrete-time double integrators, i.e., the matrices in \dref{followers} are given by
$A=\left[\begin{smallmatrix} 1 & 1\\ 0 & 1\end{smallmatrix}\right]$, $B=\left[\begin{smallmatrix} 0\\ 1\end{smallmatrix}\right].$
%$$A=\begin{bmatrix} 1 & 1\\ 0 & 1\end{bmatrix}, ~B=\begin{bmatrix} 0\\ 1\end{bmatrix}.$$
The communication graph is an undirected cycle,
where the communication channels are perturbed by stochastic uncertainties satisfying Assumption 2.
The smallest and largest nonzero eigenvalues of the Laplacian matrix are 1 and 4, respectively.
%In light of \dref{concons1n}, we can derive that the largest allowable variance of the
%channel uncertainties $\sigma_{\max}^2<1.$ 
For illustration, let the variances of the uncertainties of
the six channels be 1.5. Solving
the MARE \dref{lric} with $Q=3I$ and $\breve{\delta}^2=0.9$ gives $P =10^3\times\left[\begin{smallmatrix} 0.0319  &  0.1521\\
    0.1521 &   1.4643\end{smallmatrix}\right]$. Therefore, it follows from Theorem \ref{thm1} that
the feedback gain matrix of \dref{signals} is obtained as $K=-\left[\begin{smallmatrix} 0.1038  &1.1038\end{smallmatrix}\right]$.
The initial states of the agents are chosen to be $x_1(0)=\left[\begin{smallmatrix} 1\\0\end{smallmatrix}\right]$, $x_2(0)=\left[\begin{smallmatrix} 2\\-1\end{smallmatrix}\right]$, $x_3(0)=\left[\begin{smallmatrix} -1\\0.5\end{smallmatrix}\right]$, $x_4(0)=\left[\begin{smallmatrix} 0.8\\2\end{smallmatrix}\right]$, $x_5(0)=\left[\begin{smallmatrix} 2\\3\end{smallmatrix}\right]$, and $x_6(0)=\left[\begin{smallmatrix} 0\\1\end{smallmatrix}\right]$.
{The average of 1000 sample trajectories of the relative states $x_i(k)-x_1(k)$, $i=2,\cdots,6$,
under \dref{signals} with $\alpha=0.25$ and $K$ designed as above, are depicted in Fig. \ref{fig5}, which is an approximation of the expectation of $x_i(k)-x_1(k)$ and from which it can be observed that mean square consensus is indeed achieved.}

\section{Conclusions}\label{s7}

In this paper we have studied a robust consensus problem for a
network of general discrete-time linear agents coordinating through
uncertain communication channels. We model the communication channel
as an ideal transmission system subject to a multiplicative
stochastic perturbation. We have presented sufficient robust consensus
conditions, which exhibit intrinstic limitations imposed by the
synchronizability factor, the topological entropy of the agents, and
the mean square channel capacity. Under this limit, consensus
protocols can be designed by solving an MARE.

%%%%%%%%%%%%%%%%%%%%%%%%%

\end{document}